# Gender differences in lying in sender-receiver games: A meta-analysis

Valerio Capraro

Middlesex University

London

United Kingdom

V.Capraro@mdx.ac.uk

**Abstract**

*Whether there are gender differences in lying has been largely debated in the past decade. Previous studies found mixed results. To shed light on this topic, here I report a meta-analysis of 8,728 distinct observations, collected in 65 Sender-Receiver game treatments, by 14 research groups. Following previous work and theoretical considerations, I distinguish three types of lies: black lies, that benefit the liar at a cost for another person; altruistic white lies, that benefit another person at a cost for the liar; Pareto white lies, that benefit both the liar and another person. The results show that gender differences in lying significantly depend on the consequences of lying. Specifically: (i) males are significantly more likely than females to tell black lies (N=4,161); (ii) males are significantly more likely than females to tell altruistic white (N=2,940); (iii) results are inconclusive in the case of Pareto white lies (N=1,627).*

## Introduction

Many economic and social interactions are characterized by asymmetric information. In these situations, people may be tempted to misreport their private information. Although standard economic theory predicts that people would lie as long as that is beneficial to themselves, empirical research in economics and psychology has shown that people do not always lie. Cases in which people act honestly abound, even when being dishonest would be beneficial to all parties involved (Erat & Gneezy, 2012; Cappelen, Sørensen & Tungodded, 2013; Biziou-van-Pol, Haenen, Novaro, Occhipinti-Liberman & Capraro, 2015).

Why do some people act honestly while others do not?

Previous studies have approached this question from several angles. For example, scholars have explored the role of social and moral preferences (Biziou-van-Pol et al, 2015; Levine & Schweitzer, 2014; Levine & Schweitzer, 2015; Shalvi & de Dreu, 2014; Weisel & Shalvi, 2015), the role of incentives (Dreber & Johannesson, 2008; Erat & Gneezy, 2012; Fischbacher & Föllmi-Heusi, 2013; Gneezy, 2005; Gneezy, Kajackaite & Sobel, 2018; Mazar, Amir & Ariely, 2008; Sutter, 2009), the role of group-serving lies versus individual-serving lies (Cohen, Gunia, Kim-Jun & Murnighan, 2009; Conrads, Irlenbusch, Rilke & Walkowitz, 2013; Gino, Ayal & Ariely, 2013; Wiltermuth, 2011), and the role of manipulating cognitive resources (Gino, Schweitzer, Mead & Ariely, 2011; Shalvi, Eldar & Bereby-Meyer, 2012; Gunia et al., 2012; van't Veer, Stel & van Beest, 2014; Capraro, 2017; Barcelo & Capraro, 2017; Lohse, Simon & Konrad, 2018).

Another line of research that has received a great deal of attention is whether there are gender differences in lying. An early paper by Dreber and Johannesson (2008) found that males lie more than females, at least in the domain of black lies, that is, lies that benefit the liar at a cost for another person. This result was successfully replicated in some studies (Friesen & Gangadharan, 2012; Capraro, Schulz & Rand, 2018) but not in others (Childs, 2012; Capraro & Peltola, 2018), which found no gender differences in the context of black lies. Subsequently, Erat and Gneezy (2012) observed that the sign of gender differences in lying might depend on the consequences of the lie: they found that males lie more than females in the context of Pareto white lies (lies that benefit both the liar and another person), but females lie more than males in the context of altruistic white lies (lies that benefit another person at a cost for the liar). However, the former result was not replicated by Cappelen et al (2013), who found no gender differences in the context of Pareto white lies; and the latter result was not replicated by Biziou-van-Pol et al (2015), who, in fact, found the opposite, that males tell more altruistic white lies than females. These mixed results suggest that gender differences in lying, if existent, might be small and depending on the consequences of lying. Thus, passing to a meta-analytic approach can be useful to shed light on the topic.

The contribution of this work is to do a step in this direction by analyzing a large sample of more than 8,500 observations, coming from 65 different treatment conditions, conducted by 14 different research groups, by taking also into account the consequences of lying.

**Measure of honesty**

Researchers have developed several measures of honest behavior. For example, in Fischbacher and Föllmi-Heusi (2013), participants roll a die, in private, and then report the resulting outcome knowing that they will be paid an amount equal to the number they report, unless the number is six, in which case they do not get any payment. Thus participants have an incentive to lie (unless they get a five) for their benefit. See also Greene & Paxton (2009), Fosgaard, Hansen & Piovesan (2013), Ploner & Regner, (2013), Shalvi & Leiser (2013),

Pascual-Ezama, Prelec & Dunfield (2013), van't Veer, Stel & van Beest (2014). Conceptually similar is the matrix search task (Mazar, Amir & Ariely, 2008) and the visual perception task (Gino, Norton & Ariely, 2010). The common denominator of these paradigms is that participants complete a task and then they are paid according to the self-reported performance in this task. Thus, also in this case, participants are incentivized to lie for their own benefit.

Conceptually different, instead, is the so-called *Sender-Receiver game* (also known as *Deception game*). There are various formulations of this game (Gneezy, 2005; Erat & Gneezy, 2012), differing in relatively minor details regarding the strategy space or the type of information provided. But the general structure is as follows. The experimenter gives a piece of information (for example, the outcome of a die) to Player 1, but not to Player 2. Then Player 1 is asked to report this information to Player 2. The role of Player 2 is to guess the original piece of information (for example, the true outcome of the die). If Player 2 guesses the original piece of information, then Player 1 and Player 2 get paid according to Option A; if Player 2 does not guess the original piece of information, then Player 1 and Player 2 get paid according to Option B. Only Player 1 knows the exact allocations of money corresponding to Option A and Option B. One variant of the deception game was introduced by Biziou-van-Pol et al. (2015), in order to avoid the problem of *sophisticated deception* (i.e., Player 1 telling the truth because he or she expects that Player 2 will not believe him or her, Sutter 2009). In this variant, Player 2 has no active choice: whether participants are paid according to Option A or Option B depends only on whether Player 1 decides to lie or to tell the truth.

The Sender-Receiver game is particularly interesting because it allows to distinguish four types of lies, depending on the payoffs associated to Option A and Option B. Employing the terminology introduced by Erat and Gneezy (2012), I use the following taxonomy: *black lies* are those that benefit the liar at the expenses of the other person; *altruistic white lies* are those that benefit another person at a cost for the liar; *Pareto white lies* are those that benefit both the liar and the other person; *Spiteful lies* are those that harm both the liar and the other person. The goal of this work is to study gender differences on lying as a function of the consequences of lying.

**Theoretical considerations**

As mentioned above, previous empirical work suggests that the sign of gender differences in lying may depend on the consequences of the lie. The existence of such a dependence is in fact expected and can be actually derived from considerations regarding gender differences in social preferences and deontological judgments.

Previous work shows that males are more selfish than females in the dictator game, at least among students and Mechanical Turkers (Croson & Gneezy, 2009; Branas-Garza, Capraro & Rascón-Ramírez, 2018; Rand et al, 2016), although perhaps not in the general population (Carpenter, Connolly & Myers, 2008; Cappelen, Nygaard, Sørensen & Tungodden, 2015). Several studies have also provided evidence that males donate less than females to charity (De Wit & Bekkers, 2016; Mesch et al, 2006; Piper & Schnepf, 2008). In line with this view, social role theorists argue that males are more agentic and independent, whiles females are more unselfish and communal (Eagly, 1987). These observations suggest that self-regarding motivations may push males to tell more black lies than females, even when their intrinsic costs of lying are, on average, the same.

Regarding gender differences in altruistic behavior, an influential work by Andreoni and Vesterlund (2001) found that females are more altruistic than males when the altruistic action coincides with the egalitarian action, but males are more altruistic than females when the altruistic action is socially efficient. This suggests that gender differences in the decision to tell altruistic white lies may depend on the actual consequences of lying, such that females may tell more altruistic lies than males when lying minimizes payoff differences, while males may tell more altruistic white lies than females when lying is socially efficient.

Regarding Pareto white lies, two different arguments lead to the prediction that it is likely that there are no major gender differences in lying. On the one hand, Pareto white lies – at least those studied in previous literature and reported in this paper – are both socially efficient and egalitarian. Thus, the Andreoni and Vesterlund's (2001) result mentioned above suggests that females and males might be equally motivated to tell Pareto white lies. An alternative argument descends from considerations about the morality of lying. According to deontological ethics, an action is morally good if it instantiates certain rules or ethical norms, regardless of the consequences, and is morally good if it violates them. Telling the truth when lying is beneficial to all parties involved is thus a typically deontological choice, as it corresponds to following the rule "don't lie", regardless of consequences. Therefore, the question whether there are gender differences in telling Pareto white lies can be seen as a particular specification of the more general question of whether there are gender differences in deontological judgments. Previous research suggests that females are more deontological than males, but only in moral dilemmas that involve directly harming others for the greater good (Capraro & Sippel, 2017; Fumagalli et al, 2010; Friesdorf et al, 2015). Since lying in a Pareto white lie condition does not involve direct harm to others, also this argument suggests that it is likely that there are no major gender differences in telling Pareto white lies.

## Data collection

Data collection proceeded in several steps. First, on April 1, 2016, I announced my plan of conducting a meta-analysis of two-player sender-receiver games on the ESA Experimental

Methods Discussion Google Group. In this way, the scholars interested in having their work included in the meta-analysis could send me the raw data of their experiment(s). In the days after, I have also conducted a 2x4 google scholar search looking for pairs of keywords of the shape [gender, sex] x [honesty, dishonesty, lying, deception], and I emailed the authors of all relevant papers and requested the raw data of their experiment(s). In doing so, I received raw data of 18 different experimental treatments (some published, some not), to which I have added 32 different experimental conditions of my research group (some published, some not). To minimize file-drawer effects, I included in the meta-analysis also unpublished studies. Then I wrote a first draft of the meta-analysis (6,508 observations), which I posted on SSRN on March 11, 2017. I left this first draft online for almost one year. Then, on January 25, 2018, I emailed the ESA Experimental Methods Discussion Google Group again, announcing my plan to revise the meta-analysis. In this occasion, scholars whose work had not been included in the first version of the meta-analysis could send me the raw data of their experiment(s). I sent the same email also to the Society for Judgment and Decision Making emailing list, and I made again the same 2x4 google search that I had made before. In this second version of the meta-analysis[1], I thus analyze a total of 65 experimental treatments (36 from my own research group and 29 from 13 different research groups), for a total of 8,728 *distinct* observations (4,161 in *black lies* conditions, 2,940 in *altruistic white lies* conditions, 1,627 in *Pareto white lies* condition, and 0 in *spiteful lies* conditions). *Distinct* means that, in case a subject participated in more than one study (some studies were conducted on Amazon Mechanical Turk, so I could keep track of subjects using their MTurk ID and their IP address), I keep only the first observation. Similarly, in case the data come from iterated games, I keep only the first observation.

[1]After receiving the comments from the referees and while preparing the revision, I have emailed the ESA Experimental Methods Discussion Group and the Society for Judgment and Decision Making emailing list again to collect more data. However, I was not sent any more data this time.

**Overall analysis**

I start analyzing all 65 studies together. To do so, for each single study, I use logit regression to compute the effect of gender on honesty (which is a binary variable) with and without control on age and level of education (when known)[2]. Logit regression applied to single studies has the limitation that, if the dependent variable can be perfectly predicted, it returns no coefficient. For example, if, for a given study, all females act honestly, then logit regression returns no coefficient. This happens rarely in these dataset (3 studies over 65). When this happens, I do a correction by adding one data point by hand in such a way to maintain the sign of the effect.

Then I build a .csv file with thirteen columns: *study*, *genderc*, *genderse*, *genderc_control*, *genderse_control*, *altruistic_lie*, *black_lie*, *capraro, levine, greenberg, kouchaki, rode, cohen*, where, for each study, *genderc* (resp. *genderse*) is the coefficient (resp. the standard error) of the logit regression predicting honesty as a function of gender without control on age and level of education; similarly, *genderc_control* (resp. *genderse_control*) is the coefficient (resp. the standard error) of the logit regression predicting honesty as a function of gender with control on age and level of education*;* *altruistic_lie* and *black_lie* are two dummy variables that represent the consequences of lying in the corresponding sender-receiver game[3];

[2]I include controls for age and education because they seem to have an effect on lying (logit regression over the pool of studies; age: coeff=0.005, z=2.40, p=0.016; education: coeff=-0.081, z=-3.90, p<.001). Moreover, age has a significant effect on gender (coeff =0.116, z=5.94, p<.001), while education has no effect on gender (coeff=-0.019, z=-0.92, p=0.357).

[3]I do not include a dummy variable to represent the actual consequences of lying in the case of altruistic white lies (socially efficient vs egalitarian) because in all studies analyzed in this work lying in the altruistic white lie conditions is always socially efficient.

and *capraro, levine, greenberg, kouchaki, rode, cohen* are dummy variables representing the research group of the corresponding study, which I include as a potential moderator[4].

On this new .csv file, I first conduct a meta-regression by launching the stata command *metareg genderc altruistic_lie black_lie capraro levine greenberg kouchaki rode cohen , wsse(genderse)* and then I test whether the gender effect depends on the categorical variable of lie type by launching the command *test altruistic_lie black_lie*. In doing so, I find a significant effect (p = 0.03) of lie type, suggesting that, indeed, gender differences in lying depends on the consequences of the lie.

Subsequently, I test for the effect of research group by launching the command *test capraro levine greenberg kouchaki rode cohen*. In doing so, I find no significant effect (p = 0.506). Thus, gender differences in lying do not depend on the research group that conducted the experiment.

Now, to look at the effect of gender on lying, I conduct random-effect meta-analysis by launching the command: *metan genderc genderse, random label(namevar=study).* The results, reported in Figure 1, clearly show a significant overall effect such that females are more honest than males (effect size = 0.271, 95% CI = [0.172,0.370], Z = 5.37, p < 0.001). This effect is robust after controlling for age and level of education (effect size = 0.283, 95% CI = [0.182,0.384], Z = 5.49, p < 0.001). Furthermore, there is no evidence of heterogeneity across studies in the true size of this effect (without control: p = 0.553; with control: p = 0.631).

---

[4]I include as moderators only those research groups for which I have more than one study.

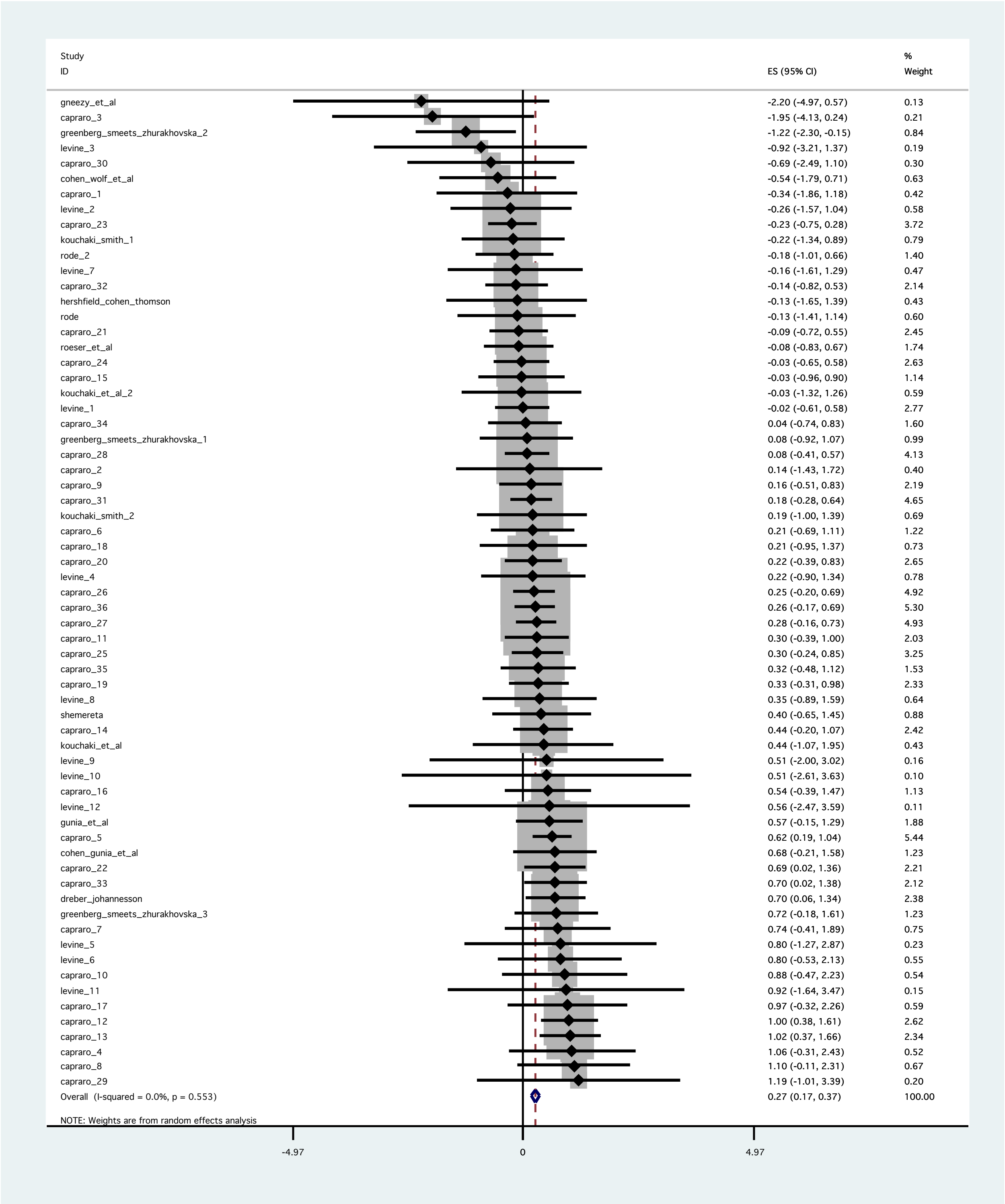

*Figure 1. Meta-analysis of gender differences on lying across all 65 studies (with no control on age and level of education).*

Finally, to test for potential publication bias and small study effect, without and with control on age and education, I conduct Egger's test and Begg's test by launching, respectively, the *Stata* commands: *metabias genderc genderse, egger*; *metabias genderc_control genderse_control, egger*; *metabias genderc genderse, begg*; and *metabias genderc_control genderse_control, begg*. In doing so, I find no evidence of publication bias and small study effect (Egger's test: without control: t = -0.99, p = 0.328; with control: t = -0.67, p=0.507; Begg's test: without control: z = -0.62, p=0.533; with control: z = -0.44, p=0.659).

## Black lies

I now analyze gender differences on the decision to tell black lies, i.e., lies that benefit the liar at the expenses of another person.

### Dataset

I analyze N = 4,161 distinct observations, coming from 36 different experimental conditions: fourteen conducted by my research group, six by Emma Levine (unpublished), one by Sheremeta & Shields (2013), three by Greenberg, Smeets, and Zhurakhovska (2015), one by Dreber and Johannesson (2008), one by Cohen, Gunia, Kim-Jun and Murnighan (2009), one by Cohen, Wolf, Panter and Insko (2011), one by Gunia et al (2012), one by Hershfield, Cohen and Thomson (2012), two by Kouchaki and Smith (2014), two by Kouchaki, Smith-Crowe, Brief and Sousa (2013), two by Rode (2010), and one by Roeser et al (2016).

### Analysis

On average, 36% of males versus 44% of females are honest. Random-effects meta-analysis shows that females are significantly more honest than males (effect size = 0.191, 95% CI = [0.058,0.325], Z = 2.81, p = 0.005). This effect is robust after controlling for age and, when possible, for level of education (effect size = 0.189, 95% CI = [ 0.054,0.325], Z = 2.74, p =

0.006). Furthermore, there is no evidence of heterogeneity across studies in the true size of this effect (without control: p = 0.858; with control: p = 0.891). Forest plot of the meta-analysis is reported in Figure 2. Finally, Egger's test (without control: z = -0.21, p = 0.833; with control: z = 0.42, p=0.681) and Begg's test (without control: z = -0.22, p=0.824; with control: z = 0.41, p=0.682) show no evidence of publication bias and small study effect.

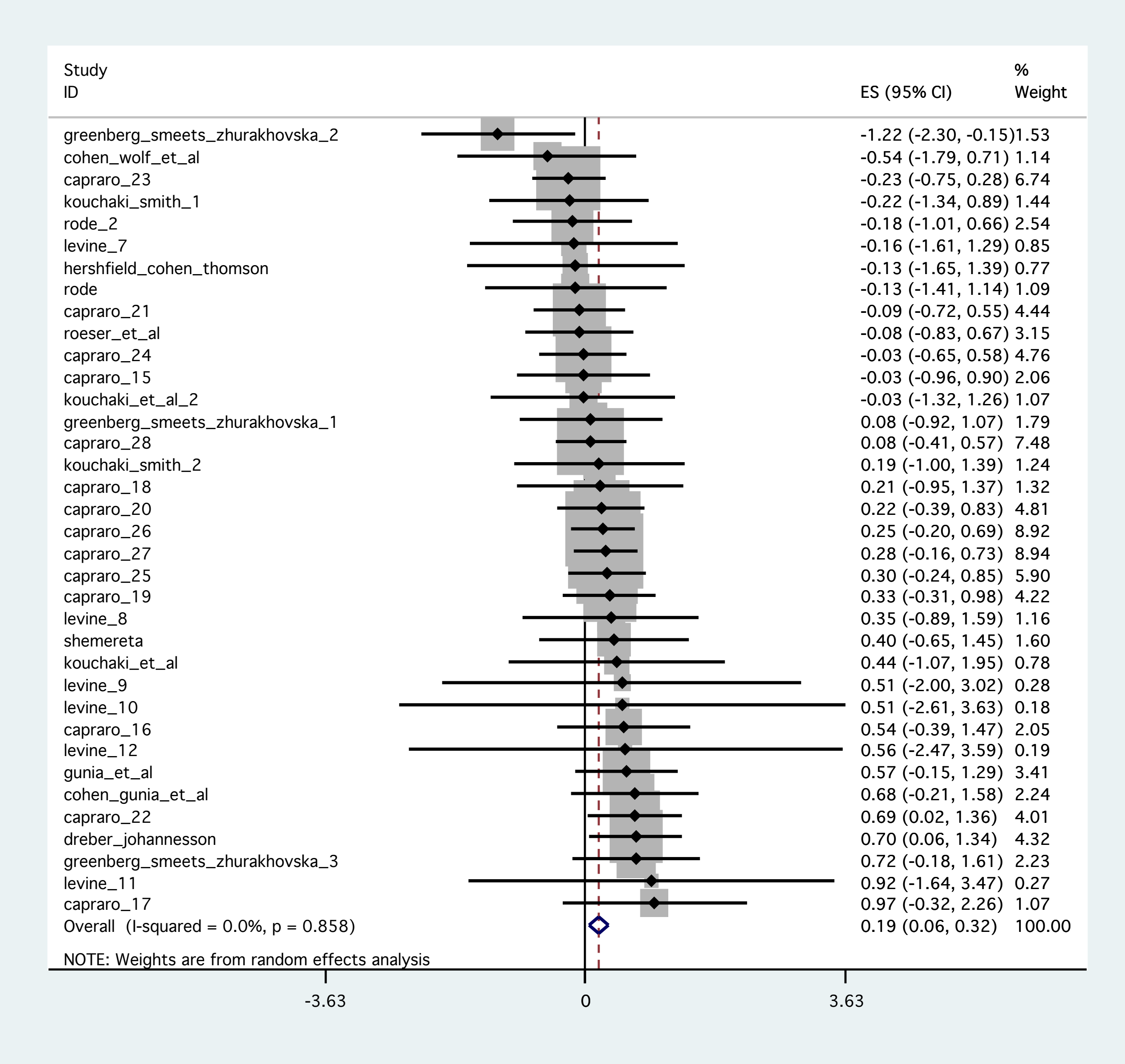

*Figure 2. Forest plot of the meta-analysis of the gender differences in telling black lies (with no control on age and level of education).*

It is formative to note that the overall effect is relatively small, and this might explain why previous research failed to consistently detect gender differences. A power analysis indeed shows that to detect the overall effect with power 0.9 at a 5% significant level one needs a sample of size N=1,989.

## Altruistic white lies

Next I analyze gender differences on the decision to tell altruistic white lies, namely, lies that benefit another person at a cost for the liar.

### Dataset

I analyze N = 2,940 distinct observations, in 20 experimental conditions: fourteen by my research group, and six (unpublished) by Emma Levine's. In all these conditions, lying is socially efficient, whereas telling the truth minimizes payoff differences.

### Analysis

On average, 76% of males versus 83% of females acts honestly. Random-effects meta-analysis finds that females are more honest than males (effect size = 0.469, 95% CI [0.256,0.681], $Z = 4.33$, $p < 0.001$). This effect is also robust after controlling for sex and, when possible, for the level of education (effect size = 0.537, 95% CI [0.341,0.733], $Z = 5.37$, $p < 0.001$). Furthermore, there is no evidence of heterogeneity across studies in the true size of this effect (without control: $p = 0.315$; with control: $p = 0.493$). Forest plot of the meta-analysis is reported in Figure 2. Finally, there is no evidence of publication bias and small studies effect (without control: Egger's test: $t = -0.70$, $p = 0.494$; Begg's test: $z = -0.59$, $p = 0.552$. With control: Egger's test: $t = -0.94$, $p = 0.359$; Begg's test: $z = -1.36$, $p = 0.172$).

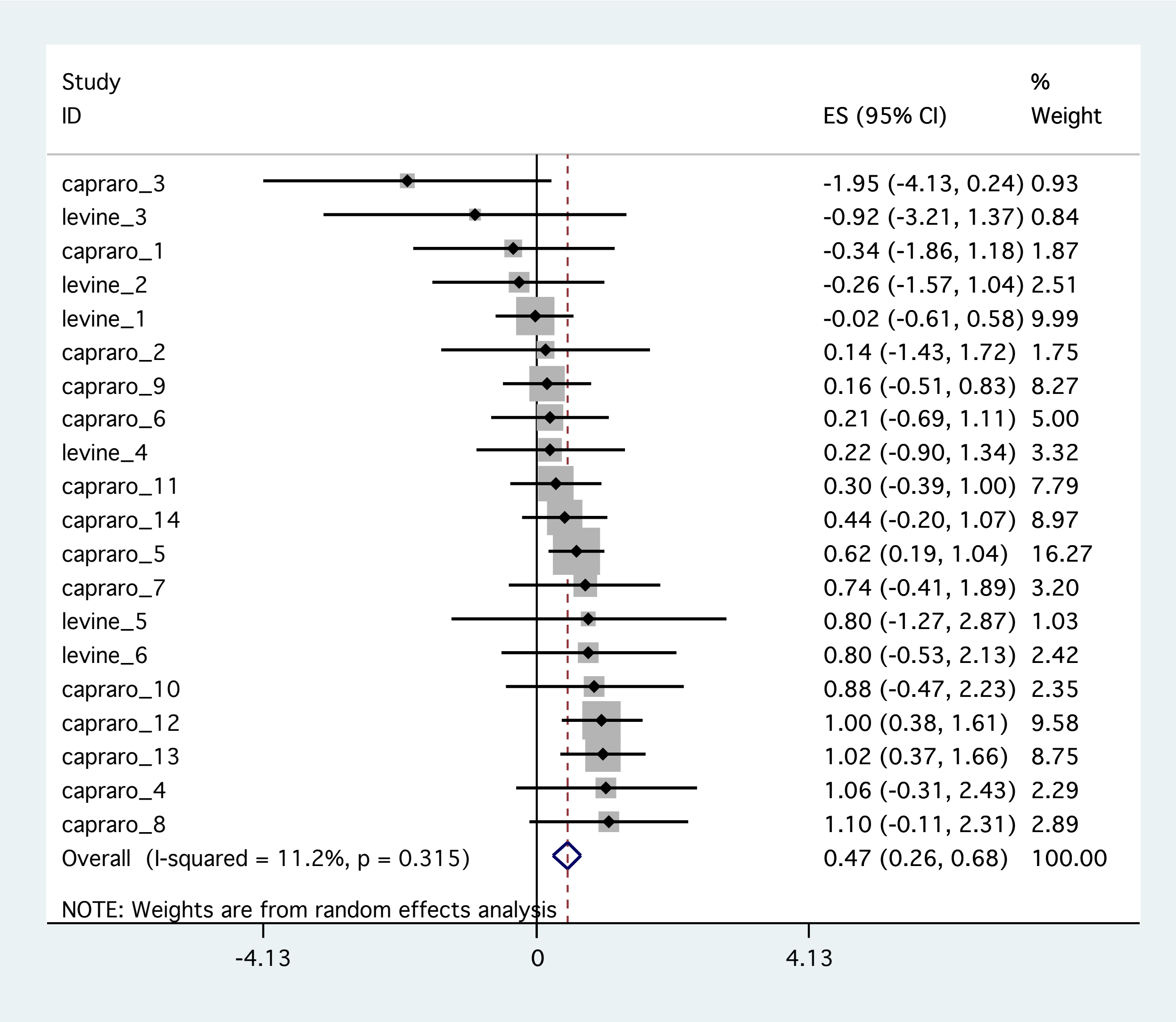

*Figure 3. Forest plot of the meta-analysis of the gender differences in telling Altruistic white lies (with no control on age and level of education).*

Also in this case, it is formative to note that the overall effect is relatively small, and this might explain why previous research failed to consistently detect gender differences. A power analysis indeed shows that to detect the overall effect with power 0.9 at a 5% significant level one needs a sample of size N=728.

## Pareto white lies

Finally, I explore gender differences on the decision to tell Pareto white lies, that is, lies that benefit both the liar and another person.

### Dataset

I analyze N = 1,627 distinct observations, in 9 experimental conditions: eight by my research group, and one by Gneezy, Rockenbach and Serra-Garcia (2013).

### Analysis

On average, 26% of males versus 29% of females acts honestly. Random-effect meta-analysis finds that males are almost significantly more dishonest than females, when I do not control for age and level of education (effect size = 0.205, 95% CI = [-0.029, 0.439], Z=1.72, p=0.085). However, this almost significant effect is not robust after controlling for age and level of education (effect size = 0.199, 95% CI = [-0.041, 0.438], Z=1.62, p=0.104). There is no heterogeneity across studies (without control: p = 0.438; with control: 0.505) and no evidence of publication bias, neither without control (Egger's test: t = -0.98, p = 0.360; Begg's test: z = -0.63, p = 0.532), nor with control on age and education (Egger's test: t = -0.84, p = 0.430; Begg's test: z = -0.63, p = 0.532).

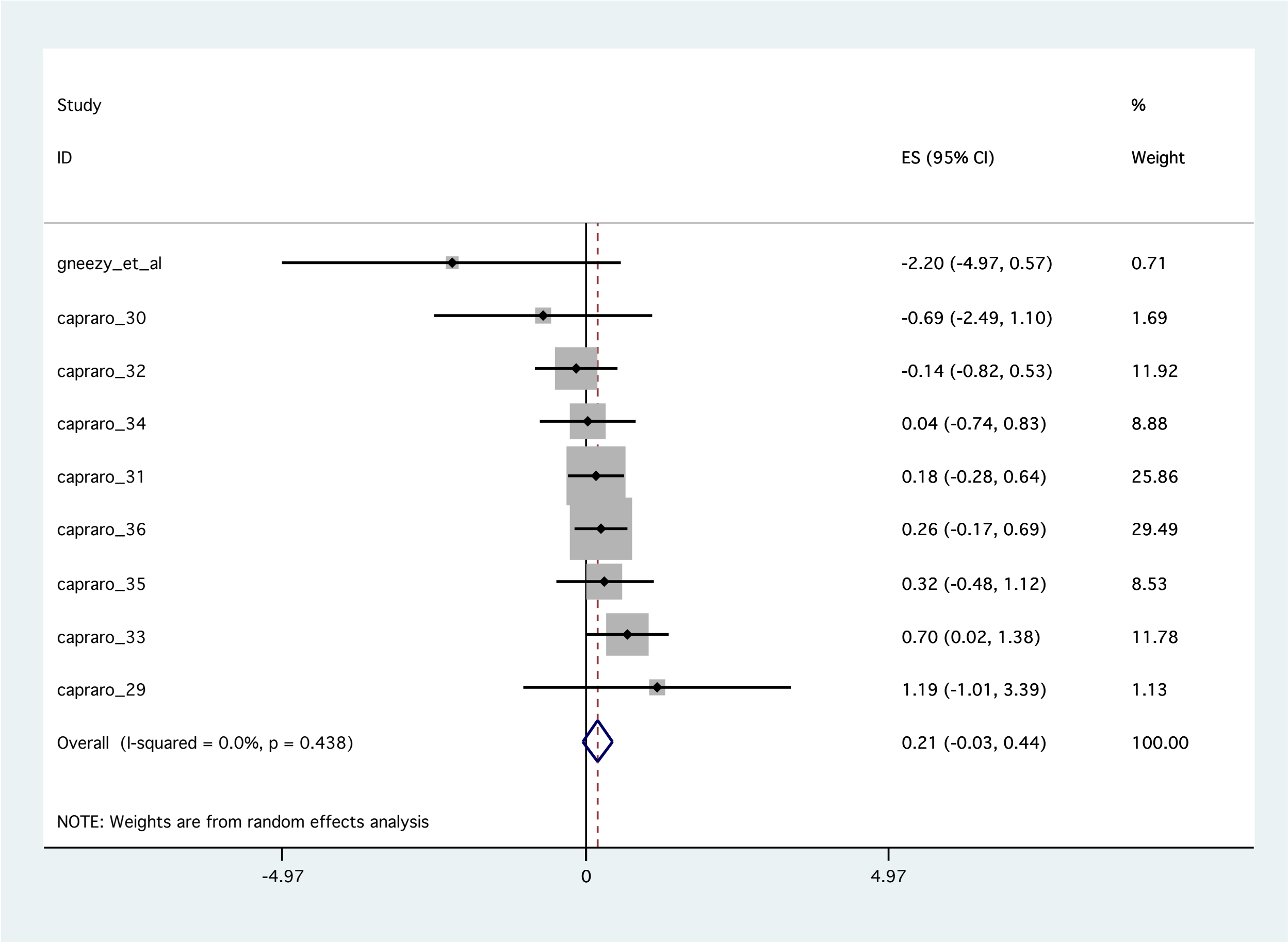

*Figure 3. Forest plot of the meta-analysis of the gender differences in telling Pareto white lies (with no control on age and level of education).*

## Discussion

In this work, I have analyzed gender differences in lying using a dataset of 8,728 distinct observations, collected using the sender-receiver game, in 65 experimental treatments, from 14 research groups. Following previous work and motivated by theoretical considerations, I have distinguished three types of lies: black lies, altruistic white lies, and Pareto white lies. The results show that gender differences in lying significantly depend on the consequences of the lie. Specifically, I have found that: (i) males are significantly more likely than females to tell black lies; (ii) males are significantly more likely than females to tell altruistic white lies; (iii) results are inconclusive in the case of Pareto white lies.

To the best of my knowledge, this is the first meta-analysis on gender differences in lying, which also takes into account the consequences of lying. The closest work I am aware of is indeed a meta-analysis of empirical studies measuring (dis)honesty using the die-under-cup task (Abeler, Nosenzo & Raymond, in press). In the die-under-cup paradigm, subjects roll a die, privately, and then are paid according to the outcome they report. In this way subjects are incentivized to misreport the outcome (Fischbacher & Föllmi-Heusi, 2013). Thus, among the black lie, the altruistic white lie, and the Pareto white lie conditions, the one that is nearer to the die-under-cup task is the black lie condition: in both the black lie condition and the die-under-cup task the liar benefits from the lie and the lie harms someone else (although the negative effect of the lie on someone else is somewhat more salient in the sender-receiver game, where another player is directly harmed, than in the die-under-cup task, where the experimenter is indirectly harmed). In line with the current meta-study, also Abeler et al (in press) finds that males are more likely than females to lie.

The main innovation of the current work is to consider also the consequences of lying. Taking them into account is crucial, especially in light of previous work on lying aversion, social preferences, and deontological judgments, which suggest that gender differences in lying may depend on the consequences of the lie. For example, Erat and Gneezy (2012) found that females are more dishonest than males in the case of altruistic white lies, while males may be more dishonest than females in the case of Pareto white lies.

In contrast to Erat and Gneezy (2012), the current meta-analysis shows that males are more dishonest than females also in the case of altruistic white lies. However, it is important to note that the current analysis does not include the data from Erat and Gneezy (2012)[5], showing the opposite effect. One may thus wonder whether including Erat and Gneezy (2012) might change the results. To address this point, I have conducted a robustness check by estimating the logit regression coefficients in Erat and Gneezy (2012) from the results reported in their paper[6]. Re-running the meta-analysis by adding this estimated coefficient does not change the qualitative result: males still appear to lie significantly more than females (without control: effect size = 0.395, 95% CI [0.158,0.632], Z = 3.27, p = 0.001; with control: effect size = 0.450 95% CI [0.221,0.680], Z = 3.84, p < 0.001). This suggests that the original finding by Erat and Gneezy (2012) might have been a false positive.

The current results are inconclusive in the case of Pareto white lies. Males seem to be slightly more dishonest than female, but the results are almost significant without control on age and level of education, and even becomes non-significant when controlling for these demographics. However, the p-values are relatively small, suggesting that there might be a small effect that I was unable to detect due to the limited power[7]. One might thus wonder whether the effect would become significant with a larger sample. As far as I know, there is only one paper using Pareto white lies that it is not included in this meta-analysis, and this is this work by Cappelen et al (2013)[8]. Does including this study resolve this inconclusiveness? Unfortunately, this does not happen, fundamentally because Cappelen et al's (2013) results are actually trending in the opposite direction, with females slightly more likely to lie than males. More formally, by estimating the coefficient from Cappelen et al (2013)[9] and rerunning the meta-analysis, I still find that males are not significantly more dishonest than females (without control: 95% CI [-0.113,0.347], Z=1.00, p=0.316; with control: 95% CI [-0.113, 0.329], Z=0.96, p=0.337). Note that the p-values are further away from significance, essentially because, as already mentioned, Cappelen et al's (2013) results are trending in the opposite direction as the original effect.

---

[5]I asked the data by email to Uri Gneezy, who replied that "the relevant data from my papers (gender and decisions) is in the papers or online appendix". Not having found the exact data on the appendix, I used the results reported in the paper to estimate the effect as reported in the main text.

[6]Erat and Gneezy (2012) conducted an altruistic white lie condition with N=101 subjects (62 males and 39 females). Also in their experiment, lying is socially efficient while being honest is egalitarian. They found a proportion of lying of 41% among females and 27% among males. To estimate the logit regression coefficient, I assume that 16 females lie versus 17 males.

[7]Power analysis shows that to detect the overall effect with power 0.9 at a 5% level one needs a sample of size N=2,646.

[8]I emailed all three authors of the Cappelen et al (2013) paper to ask for their data, but I received no answer. For this reason, I opted for estimating their effect from the result reported in their paper and include this in this additional analysis.

As described in the Theoretical Considerations section, the overall pattern of results is consistent with the hypothesis that females and males do not differ in the intrinsic cost of lying, but they differ only on social preferences: males are more selfish than females and more concerned about social efficiency than females; while females are more concerned than males about reaching an equitable distribution of payoffs. Further research should be devoted to test this other predictions of this hypothesis. For example, it would be important to explore gender differences in telling altruistic white lies in situations in which lying minimizes payoff differences, while being honest is socially efficient and maximize the individual payoff. The aforementioned view predicts that the sign of the gender difference in this case should switch.

Future research should also explore gender differences in telling spiteful lies, that is, lies that harm both players. Unfortunately, this kind of lies has been studied very little in the literature. The only study I am aware of is by Rosaz and Villeval (2012), who found only 3.9% of lying. They did not report gender differences. Exploring gender differences in the decision to tell spiteful lies is an interesting avenue for future research.

[9]To estimate the regression coefficients, I use the results reported in Cappelen et al (2013) as follows. Lying was 65.6% among females and 61.9% among males. I assume they have 200 males and 200 females, although in reality they have 352 subjects in total.

Another interesting route for further work regards finding potential moderators. The current analysis found no heterogeneity effect in the meta-analysis. Of course, this does not imply that the gender effect is not moderated by any variable. It could simply be that I was not able to detect because of insufficient power or insufficient variance. Exploring whether the gender effect is moderated by other variables could be an interesting topic for further research.

In sum, here I studied gender differences in lying using a large dataset of 8,728 observations on the sender-receiver game. I found two clear results: males are more likely than females to tell black lies, and males are more likely than females to tell altruistic white lies (at least when lying is socially efficient). Future research should explore gender differences in the case of Pareto white lies and spiteful lies, and in the case of altruistic white lies, when lying minimizes payoff differences, while being honest is socially efficient and individually optimal.